\documentclass[11pt]{article}
\usepackage{authblk}
\usepackage{amsmath}	
\usepackage{graphicx}
\usepackage[margin=1in]{geometry}
\usepackage{setspace}
\usepackage{abstract}
\pdfoutput=1

\newcommand{\degrees}{^{\circ}}
\newcommand{\norm}[2]{\left\lVert#1\right\rVert_#2}
\newcommand{\jth}[1]{$#1$th}
\newcommand{\txtpow}[1]{{\mbox{\scriptsize{#1}}}}
\newcommand{\tinytxtpow}[1]{{\mbox{\tiny{#1}}}}

\newcommand{\JournalTitle}[1]{\textit{#1}}

\title{\vspace{-2cm}Single Pixel Polarimetric Imaging through Scattering Media}

\author[1]{Kai Ling C. Seow}
\author[2]{Peter T\"or\"ok}
\author[1]{Matthew R. Foreman*}

\affil[1]{Blackett Laboratory, Department of Physics, Imperial College London, Prince Consort Road, London, SW7 2AZ, UK}
\affil[2]{Division of Physics and Applied Physics, Nanyang Technological University, Singapore}

\affil[*]{Corresponding author: matthew.foreman@imperial.ac.uk}

\begin{document}
	
	\maketitle
	
\begin{abstract}
	Polarimetric imaging can provide valuable information about biological samples in a wide range of applications. Detrimental scattering however currently limits the imaging depth of \textit{in-vivo} imaging to $\sim$1 transport mean free path. In this work, single pixel imaging is investigated as a means of reconstructing polarimetric images through scattering media. A theoretical imaging model is presented and the recovery of the spatially resolved Mueller matrix of a hidden test object is demonstrated experimentally for scattering phantoms with thicknesses up to twice the transport mean free path. 
\end{abstract}
\vspace{0.5cm}
	
Development of quantitative techniques for measurement and monitoring of biological tissue is  vital to improving healthcare and quality of life. Significant effort has thus been made to improve the sensitivity and specificity of optical bioimaging technology. Predominantly, current methods are based on measuring optical intensity or wavelength, however, such measurements forego the additional information given by the polarisation state of light. Not only does polarisation imaging offer additional contrast mechanisms, such as study of birefringence and diattenuation of collagen networks~\cite{Chue-Sang2017},  it can also reveal the micro-structure and
composition of tissues~\cite{Mazumder2014}. In turn, such information can play a key role in diagnostics and fundamental biomedical research, for example by improving discrimination of cancerous and benign tissues~\cite{Novikova2012}, enabling detection of glaucoma~\cite{Dada2014} and facilitating study of cartilage diseases~\cite{Ellingsen2019}.

Although \emph{in-vivo} bioimaging methods are sought so as to reduce the need for invasive biopsies and histological studies, they are frequently impeded by the need to image through relatively thick layers of highly scattering tissue which scrambles the spatial and polarimetric information contained within an image~\cite{Byrnes2020}. Polarisation sensitive optical coherence tomography is a well established polarimetric imaging technique which rejects scattered light by means of coherence and polarisation gating \cite{deBoer2017}. Such methods are however typically limited to depths of a transport mean free path (TMFP), approximately 1~mm in biological tissue, due to the decrease in the ballistic intensity for thicker samples. To image deeper, a range of solutions that make use of, rather than reject, scattered light have  been proposed for intensity based imaging modalities, including wavefront shaping, full transmission matrix measurements, use of speckle correlations and single pixel imaging (see Refs.~\cite{Rotter2017,Yoon2020} for a review). 

Despite the success achieved in intensity-based imaging beyond a TMFP, little progress has been made in polarimetric modalities even though it is known that polarisation information degrades over the longer length scale of a few TMFPs~\cite{Sankaran1999b}. Recovery of the polarisation state of light focused at such depths in a scattering medium has been demonstrated using broadband wavefront shaping and used for structural imaging \cite{deAguiar2017}, however the polarimetric properties of the sample were not retrieved. Moreover, whilst full vector transmission matrix measurements have been reported \cite{Tripathi2012}, to date their use has been limited to engineering of focal fields \cite{Guan2012}. This article therefore aims to demonstrate polarimetric imaging  through scattering media at length scales longer than a TMFP for the first time. To do so a single pixel polarimetric imaging setup is used \cite{Soldevila2013,Duffin2014}, which combines sequential variation of
the illumination basis and incident polarisation state with spatial integration of the polarisation resolved output to reconstruct an image~\cite{Tajahuerce2014}.  A single pixel polarimetric imaging model and image reconstruction algorithm are first discussed, before a detailed description of a proof-of-principle experimental setup is given. Experimental results of a test object hidden behind scattering phantoms of varying thickness are then presented.

\begin{figure}[t]
	\centering
	\includegraphics[width=0.98\linewidth]{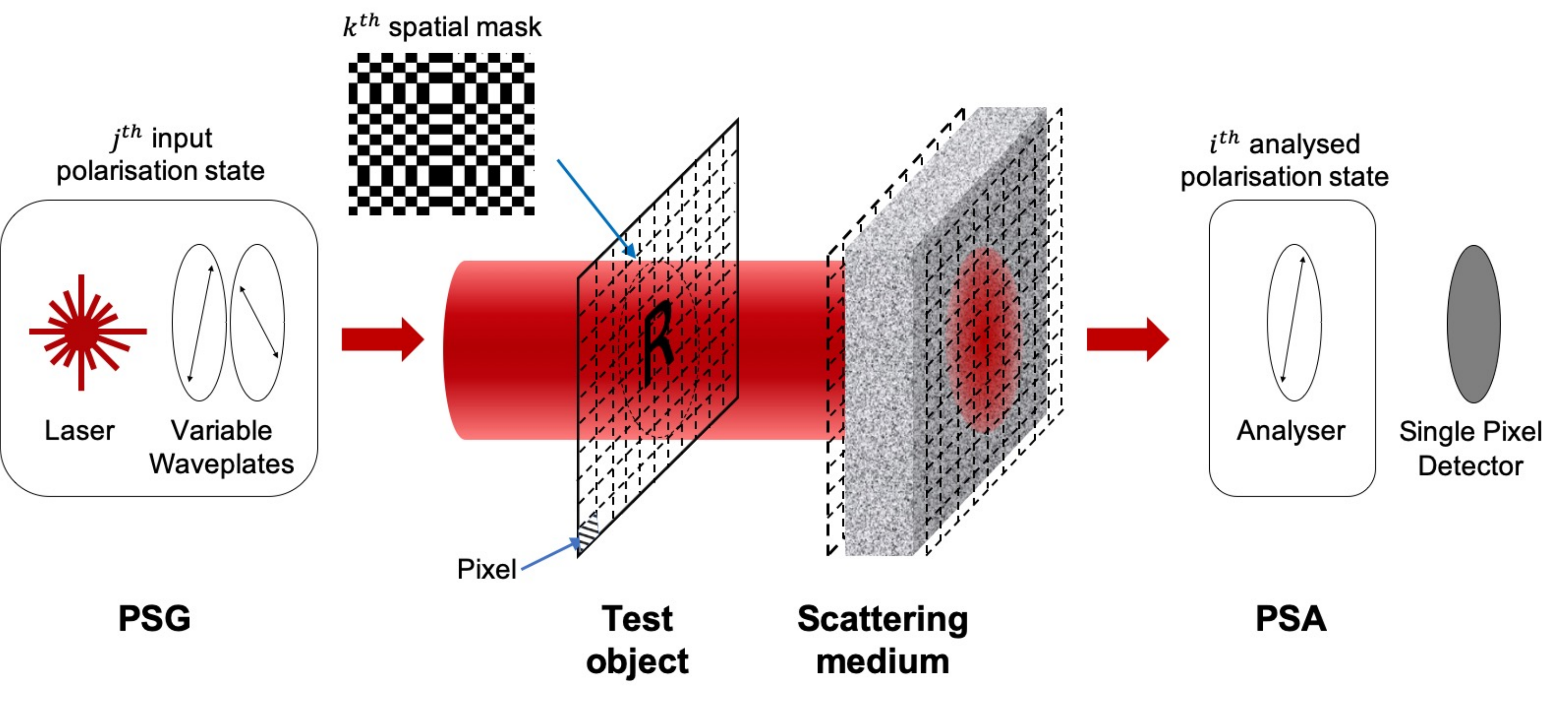}
	\caption{Schematic of a single pixel polarimetric imaging setup.}
	\label{fig:imagingconfig}
\end{figure}

The imaging configuration considered in this work is shown in Figure \ref{fig:imagingconfig}. A test object, hidden behind a static scattering medium, is illuminated by a coherent spatially modulated beam with a specific input polarisation state as generated by a polarisation state generator (PSG). Light transmitted through the object and scattering medium is then passed through a polarisation state analyser (PSA), which projects the incident light onto a test polarisation state, before it is subsequently collected by a single pixel detector which has no spatial resolution. It will be shown that the full polarimetric properties of the object, as described by its spatially dependent Mueller matrix, can then be found using multiple measurements with different input polarisation states, analysis states and illumination profiles.

To model the polarimetric imaging process consider first discretising the transverse spatial coordinates into individual pixels. The illumination field incident on the \jth{m} pixel of the object can then be described using the spatially dependent Jones vector $\vec{E}^{\txtpow{inc}}_{mj}  =  \psi^{k}_{m} \vec{E}_{j} $, where $\psi^{k}_{m}$ describes the amplitude modulation  of the \jth{k} input spatial mode and $\vec{E}_{j}$ is the Jones vector for the \jth{j} input polarisation state. Letting $\mathbf{T}^\txtpow{{obj}}_{m}$ denote the Jones matrix of the \jth{m} pixel of the object, the field at the input surface of the scattering medium is hence  $ \vec{E}^{\txtpow{obj}}_{mj} =  \mathbf{T}^{\txtpow{obj}}_{m} \vec{E}^{\txtpow{inc}}_{mj} $. Assuming any imaging optics present do not affect the polarisation state, the Jones vector in the  plane after the PSA can  be expressed as
\begin{equation}
\begin{aligned}
\vec{E}^{\txtpow{out}}_{nijk} = \mathbf{T}_{i}\sum_{m} \mathbf{T}^{\txtpow{SM}}_{nm} \vec{E}^{\txtpow{obj}}_{mj}  = \mathbf{T}_{i}\sum_{m}  \psi^{k}_{m}  \mathbf{T}^{\txtpow{SM}}_{nm} \mathbf{T}^{\txtpow{obj}}_{m} \vec{E}_{j}\ ,
\end{aligned}
\label{SPPI:eq9}
\end{equation}
where the $2 \times 2$ Jones matrix, $ \mathbf{T}^{\txtpow{SM}}_{nm}$, relates the Jones vectors at the \jth{m} input and \jth{n} output pixels and $\mathbf{T}_{i}$ is the spatially homogenous Jones matrix of the \jth{i} PSA.

\noindent

Since the intensity measured by the single pixel detector is an incoherent sum of the contributions from all output pixels, it is convenient to use the coherency vector representation of light whereby $\vec{C} =  \vec{E} \otimes  \vec{E}^{*}$ ~\cite{bashara1977ellipsometry}. In particular, the total spatially integrated coherency vector $\vec{C}^{\txtpow{tot}}_{ijk} = \sum_{n} \vec{C}^{\txtpow{out}}_{nijk} $ is given by
\begin{equation}
\begin{aligned}
\vec{C}^{\txtpow{tot}}_{ijk	}				&=\sum_{m}  \left(\mathbf{T}_{i} \otimes \mathbf{T}_{i}^{*}\right) \mathbf{A}_{m} \vec{C}^{\txtpow{obj}}_{mjj}  + \sum_{m} \sum_{l \neq m}  \left(\mathbf{T}_{i} \otimes \mathbf{T}_{i}^{*}\right) 
\mathbf{B}_{ml} \vec{C}^{\txtpow{obj}}_{mlj}
\end{aligned}
\label{SPPI:eq11}
\end{equation}
where $\mathbf{A}_{m} = \mathbf{B}_{mm}$, $\mathbf{B}_{ml} = \sum_{n} (\mathbf{T}^{\txtpow{SM}}_{nm} \otimes \mathbf{T}^{\txtpow{SM},*}_{nl})$, 
$\otimes$ denotes the direct product and $*$ represents complex conjugation. Note that no temporal averaging is required since the illumination is coherent and all optical elements are static~\cite{bashara1977ellipsometry}. \eqref{SPPI:eq11} shows that the measured coherency vector can be split into two components. The first term is an incoherent sum of the contributions from each input pixel, while the second term describes a mixed contribution from different input pixels. In particular, noting that  elements in $\mathbf{T}^{\txtpow{SM}}_{nm}$ relate the field components for the \jth{m} input and \jth{n} output pixels, and further making an ergodic assumption whereby spatial averages are equivalent to ensemble averages, the elements in $\mathbf{B}_{ml}$  can be seen to be an estimate of the correlation between polarised fields originating from different pixels before the scattering medium.  Typically this correlation decreases as the separation between the source pixels increases, over a length scale $\kappa$ which is determined by the smallest of the translation correlation length~\cite{Judkewitz2015} or the average speckle size~\cite{Seow2020}. As such, when pixels of size larger than $\kappa$ are used, the field from each input pixel gives rise to an uncorrelated output speckle pattern. Accordingly, elements of $\mathbf{A}_{m}$ are much larger in magnitude than $\mathbf{B}_{ml}$, whereby $\vec{C}^{\txtpow{tot}}_{ijk} \approx \sum_{m} \left(\mathbf{T}_{i} \otimes \mathbf{T}_{i}^{*}\right) \mathbf{A}_{m} \vec{C}^{\txtpow{obj}}_{mjj}$.
With a sufficiently large pixel size, the total integrated Stokes vector at the single pixel detector is hence given by 
\begin{equation}
\begin{aligned}
\vec{S}^{\txtpow{tot}}_{ijk}  &\approx \mathbf{M}_{i} \sum_{m}  \mathbf{M}^{\txtpow{SM}}_{m} \vec{S}^{\txtpow{obj}}_{mjj} = \mathbf{M}_{i} \sum_{m} |\psi^{k}_{m}|^{2} \mathbf{M}^{\txtpow{SM}}_{m} \mathbf{M}^{\txtpow{obj}}_{m} \vec{S}_{j}   \ ,
\end{aligned}
\label{SPPI:eq15}
\end{equation}
where we have used the standard matrix $\mathbf{\Gamma}$ to convert between coherency vectors and Stokes vectors viz. $\vec{S} = \mathbf{\Gamma} \vec{C}$, and between Jones and Mueller matrices: $\mathbf{M} = \mathbf{\Gamma}\left( \mathbf{T} \otimes \mathbf{T}^{*} \right) \mathbf{\Gamma}^{-1}$~\cite{bashara1977ellipsometry}. Note $\vec{S}_{j}  $ is the spatially uniform Stokes vector corresponding to the \jth{j} incident Jones vector $\vec{E}_j$ and that  $\mathbf{M}^{\txtpow{SM}}_{m} =\mathbf{\Gamma}\mathbf{A}_{m}\mathbf{\Gamma}^{-1}$. 

By definition, the intensity collected by the single pixel detector, $I^{\txtpow{tot}}_{ijk}$, is given by the first element of $\vec{S}^{\txtpow{tot}}_{ijk}$, or explicitly 
\begin{equation}
\begin{aligned}
I^{\txtpow{tot}}_{ijk} = \sum_{m} | \psi^{k}_{m} | ^{2} \left( \vec{a}_{i}^{T} \mathbf{M}^{\txtpow{SM}}_{m} \mathbf{M}^{\txtpow{obj}}_{m} \vec{S}^{}_{j} \right)  = \vec{\Psi}_{k}  \cdot \vec{d}_{ij}\ ,
\end{aligned}
\label{SPPI:eq17}
\end{equation}
where $^T$ denotes transposition and the \jth{m} element of the vectors $\vec{\Psi}_{k}$ and $\vec{d}_{ij}$ correspond to $ | \psi^{k}_{m} | ^{2} $ and $( \vec{a}_{i}^{T} \mathbf{M}^{\txtpow{SM}}_{m}  \mathbf{M}^{\txtpow{obj}}_{m} \vec{S}^{}_{j} )$ respectively. The vector $\vec{a}_{i}$ is the first row of $\mathbf{M}_{i}$ and corresponds to the Stokes vector of the \jth{i} analysed polarisation state. For each input and analysed polarisation state, the collected intensity is thus seen to be a scalar projection of $\vec{d}_{ij}$ on the spatial mask, $\vec{\Psi}_{k}$. As such, by sequentially projecting spatial masks such that the vectors $\vec{\Psi}_{k}$ make up a complete spatial basis, $\vec{d}_{ij}$ can be retrieved as 
\begin{equation}
\begin{aligned}
\vec{d}_{ij} = \mathbf{\Psi}^{-1} \vec{I}^{\txtpow{tot}}_{ij}\ ,
\end{aligned}
\label{SPIinv}
\end{equation}
where $\vec{I}^{\txtpow{tot}}_{ij} = [I^{\txtpow{tot}}_{ij1},I^{\txtpow{tot}}_{ij2},\ldots]^T$  and $\vec{\Psi}_{k}$ is the \jth{k} row of the matrix $\mathbf{\Psi}$. Once $\vec{d}_{ij}$ is obtained for all input and analysed polarisation states, the set of intensity values for the \jth{m} input pixel can be related to the Mueller matrix of the test object, $\mathbf{M}^{\txtpow{obj}}_{m}$, according to $\mathbf{D}_{m} = \mathbf{AM}^{\txtpow{SM}}_{m}\mathbf{M}^{\txtpow{obj}}_{m}\mathbf{W}$, where the \jth{m} element of $\vec{d}_{ij}$ forms the \jth{(i,j)} element of $\mathbf{D}_{m}$, and the rows (columns) of the so-called instrument matrix $\mathbf{A}$ ($\mathbf{W}$) correspond to the Stokes vectors of the analysed (input) polarisation states, i.e. $\vec{a}_{i}$ ($\vec{S}_{j}$). To uniquely determine the 16 elements in the $4\times4$ Mueller matrix, $\mathbf{M}^{\txtpow{obj}}_{m}$, at least four input and analysed polarisation states are required. With suitable PSG and PSA architectures and a known $\mathbf{M}^{\txtpow{SM}}_{m}$, the spatially resolved Mueller matrix of the object, $\mathbf{M}^{\txtpow{obj}}_{m}$, can then be computed on a pixel-wise basis as $	\mathbf{M}^{\txtpow{obj}}_{m} = (\mathbf{AM}^{\txtpow{SM}}_{m})^{-1}\mathbf{D}_{m} \mathbf{W}^{-1}$. 
In practice, however, the presence of noise means such an inversion typically yields unphysical Mueller matrices. As such, in this work the Mueller matrix of the test object was instead computed using a least squares algorithm that solves for
\begin{equation}
{\mathbf{M}}_{m}^{\tinytxtpow{obj}}=\underset{\mathbf{M}}{\mbox{argmin}} \norm{\mathbf{D}_{m}-\mathbf{AM}^{\txtpow{SM}}_{m}\mathbf{M}\mathbf{W}}{2} 
\label{CLSQR_L2norm}
\end{equation} 
subject to the constraint that the related $\mathbf{H}$ matrix is positive semi-definite~\cite{Gil2000}.  In combination, \eqref{SPIinv} and \eqref{CLSQR_L2norm} allow the spatially resolved Mueller matrix of the object to be retrieved. 

The need to know $\mathbf{M}^{\txtpow{SM}}_{m}$, i.e. to pre-calibrate the scattering medium, contrasts with conventional intensity based single pixel imaging \cite{Tajahuerce2014}. Fundamentally, this difference arises since the scattering medium can change the polarisation of transmitted light such that the total transmittance for each polarisation channel differs, whereas for conventional single pixel setups the total transmitted intensity (i.e. the measurand) is a fixed proportion of the incident intensity for all measurements. Two factors, however, can help to mitigate the burden of calibration of $\mathbf{M}^{\txtpow{SM}}_{m}$. Firstly, many typical scattering media only introduce an effective depolarisation of incident light. Determination of the corresponding Mueller matrix then reduces to establishing the corresponding depolarisation lengths~\cite{Sankaran1999b} and medium thickness which is simpler then a complete Mueller matrix measurement. Secondly, for a statistically homogeneous scattering medium, an ergodic assumption can be made such that spatial averaging over a sufficiently large input pixel implies that the measured Mueller matrix is an approximation of an ensemble averaged Mueller matrix that is, consequently, independent of input pixel location~\cite{Seow2020}. As such only a single polarimetric measurement needs to be taken to determine $\mathbf{M}^{\txtpow{SM}}_{m}$ for all $m$. The latter approach is taken in this work.

Using the discussed imaging model, single pixel polarimetric imaging through scattering media was experimentally tested using custom-made scattering phantoms made from 1~$\mu$m diameter silica microspheres (\textit{Merck, Monospher 1000E}) embedded in epoxy resin (\textit{Easy Composites GlassCast 50 Clear Epoxy Casting Resin}). The fabrication procedure followed closely that discussed by Tahir \textit{et al.}~\cite{Tahir2005}. Biological tissues typically exhibit scattering anisotropy factors close to 1 and mean free paths (MFPs) $\sim 100$~$\mu$m~\cite{Ntziachristos2010}. As such, the scattering phantoms were designed to have similar scattering parameters.  Taking the refractive indices of the microspheres and cured epoxy resin to be $1.457$ and 1.55 (measured using a \textit{Bellingham \& Stanley, Abbe 5 Refractometer}) respectively, the scattering anisotropy factor of the microspheres was found using Mie theory to be $g=0.95$. The MFP of the fabricated scattering media was experimentally determined to be $l=395$~$\mu$m by fitting the measured intensity of transmitted ballistic light for scattering media of different thicknesses to the exponential decrease predicted by the Beer-Lambert law. The corresponding TMFP is $l_{tr}= l/(1-g)= 5$~mm. For the experiments reported here, three scattering media, henceforth referred to as SM1, SM2 and SM3, with $L/l = 18.57, 24.56, 43.13$ respectively ($R = L/l_{tr}=  0.85	,1.12  , 1.97$), were used.

The experimental setup used for single pixel polarimetric imaging followed the structure of Figure~\ref{fig:imagingconfig}. The PSG consisted of a laser beam with a wavelength of 638~nm (\textit{Cobolt, MLD638}) that was passed through a Glan-Thompson prism with its transmission axis oriented in the \textit{y} direction, followed by two variable waveplates (\textit{ArcOptix}) oriented at $27\pm1\degrees$ and $72\pm1\degrees$. Four input polarisation states were generated consecutively by setting the applied phase shifts as $(3\pi/4, 3\pi/4)$, $(3\pi/4, 7\pi/4)$, $(7\pi/4, 3\pi/4)$ and $(7\pi/4, 7\pi/4)$. Theoretically, this configuration minimises the condition number of $\mathbf{W}$~\cite{DeMartino2003a}, thus reducing noise amplification in the reconstruction algorithm. The beam was then spatially filtered and expanded, before it was incident on a digital micromirror device (DMD). The DMD (\textit{Texas Instruments, DLP4500}) spatially modulated the beam and was imaged onto the object plane resulting in an effective pixel size of 0.2~mm at the object plane. This pixel size was chosen to be larger than the average speckle size of the intensity speckle transmitted by SM1 thereby satisfying the pixel size requirements discussed above for all  phantoms. The object plane was then imaged onto the PSA by a 0.05 numerical aperture lens. Note that the numerical aperture of the lens affects the measurement SNR but not the imaging resolution. When a scattering medium is present, it is placed between the test object and the PSA, such that it is the light transmitted through the scattering medium that is collected. A division of amplitude PSA, analysing linearly polarised light at \textit{x}, \textit{y} and $45\degrees$ orientations, as well as left circularly polarised light, was used. The corresponding theoretical condition number of $\mathbf{A}$ is thus 3.23. Although PSA configurations with lower condition numbers are  possible~\cite{Foreman2015}, the chosen setup can be built economically using off-the-shelf components. To enhance the signal to noise ratio (SNR), lock-in detection was also implemented by modulating the intensity of the laser source using a frequency generator (\textit{TTi, TG330}) and sequentially forwarding the measured signal from the four detectors into a lock-in amplifier (\textit{Stanford Research Systems, Model SR530}). 

\begin{figure}[t]
	\centering
	\includegraphics[width=0.7\linewidth]{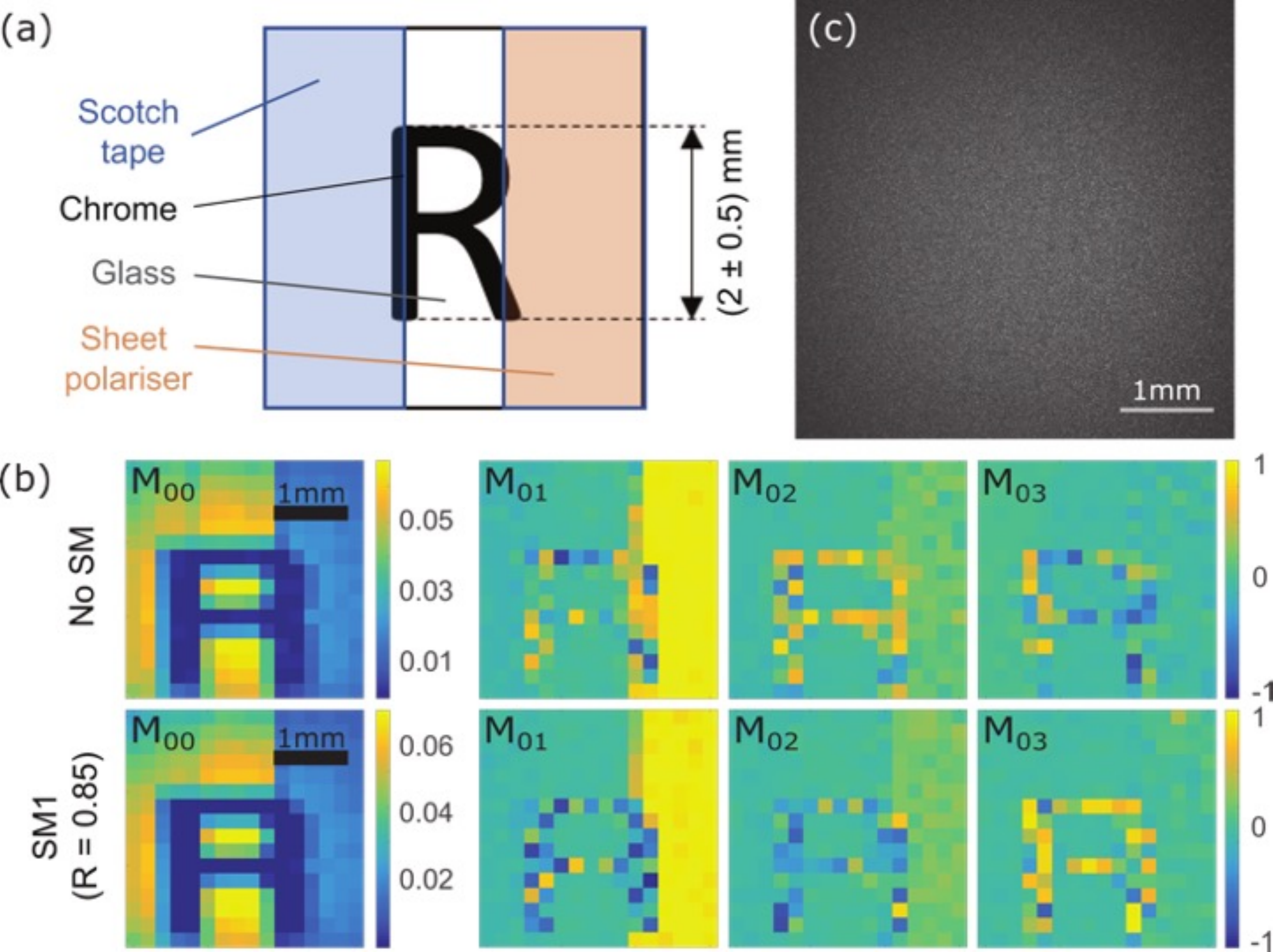}
	\caption{(a) Illustration of the test object used in the experiments.  (b) Comparison of the first row of the spatially resolved Mueller matrix obtained with and without SM1 present.  (c) Image taken by a CMOS camera.}
	\label{fig:Fig2}
\end{figure}
An illustration of the test object used in the experiments is shown in Figure \ref{fig:Fig2}(a). It consisted of a letter R  printed on a soda lime glass substrate using low-reflectivity chrome (\textit{Thorlabs, Multi-Frequency Grid Distortion Target R1L3S3P}) with a sheet polariser (\textit{Thorlabs, LPVISE2$\times$2}) and scotch tape adhered to distinct regions. The transmission axis of the sheet polariser was oriented in the $x$ direction. This test object possesses both a spatial variation in polarimetric properties (i.e. polariser, glass and retarder) as well as transmittance (i.e. the opaque letter R). Before any measurements were made, the instrument matrices, $\mathbf{A}$ and $\mathbf{W}$ were  obtained by calibrating the setup using the eigenvalue calibration method~\cite{Compain1999}. A single measurement of $\mathbf{M}_m^{\txtpow{SM}}$ was subsequently taken for each scattering medium without the test object present. Specifically, the entire object was uniformly illuminated and measurements taken for each input and analysed polarisation state. Using the known instrument matrices $\mathbf{M}_m^{\txtpow{SM}}$ was found using a constrained minimum least squares algorithm analogous to that discussed above (cf. \eqref{CLSQR_L2norm}).

Upon insertion of the test object, image data was acquired by sequentially displaying spatial masks from a scrambled Hadamard basis of order 16~\cite{Do2012} on the DMD for each input polarisation state. The corresponding intensities recorded by the photodiodes were processed for each scattering medium using \eqref{SPIinv} and \eqref{CLSQR_L2norm} to recover $\mathbf{M}^{\txtpow{obj}}_{m}$. The reconstructed image for imaging through SM1 is shown in Figure \ref{fig:Fig2}(b). For brevity, only the first row of the spatially resolved Mueller matrix is presented. The Mueller matrix measured  without any scattering medium is also shown for comparison.  The full Mueller matrix can be found in the supplementary figures. The $M_{00}$ element is presented in its original form to highlight the reconstruction of the object's unpolarised intensity transmittance, however the remaining elements are normalised by their respective $M_{00}$ values to allow for easier comparison of the polarimetric properties of each pixel.  Qualitatively, it can be seen that the Mueller matrix obtained with and without  SM1 present are very similar. In contrast, an image taken with a CMOS camera (Figure \ref{fig:Fig2}(c)) exhibits a speckle pattern with no correspondence to the test object.

\begin{figure}[t]
	\centering
	\includegraphics[width=0.7\linewidth]{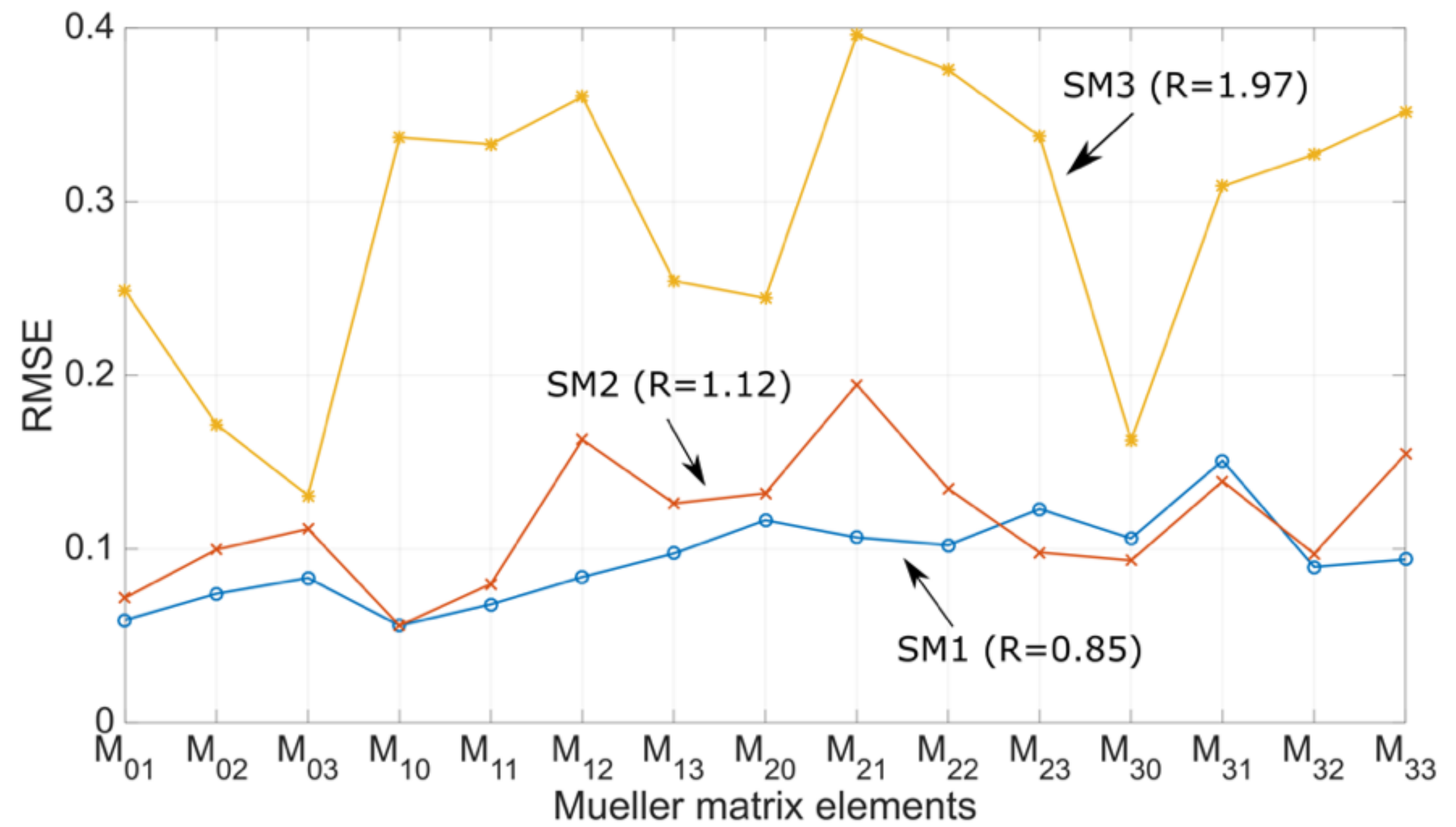}
	\caption{RMSE for the normalised Mueller matrix elements for each scattering phantom.}
	\label{fig:RMSE}
\end{figure}
\noindent

\begin{figure}[t]
	\centering
	\includegraphics[width=0.7\linewidth]{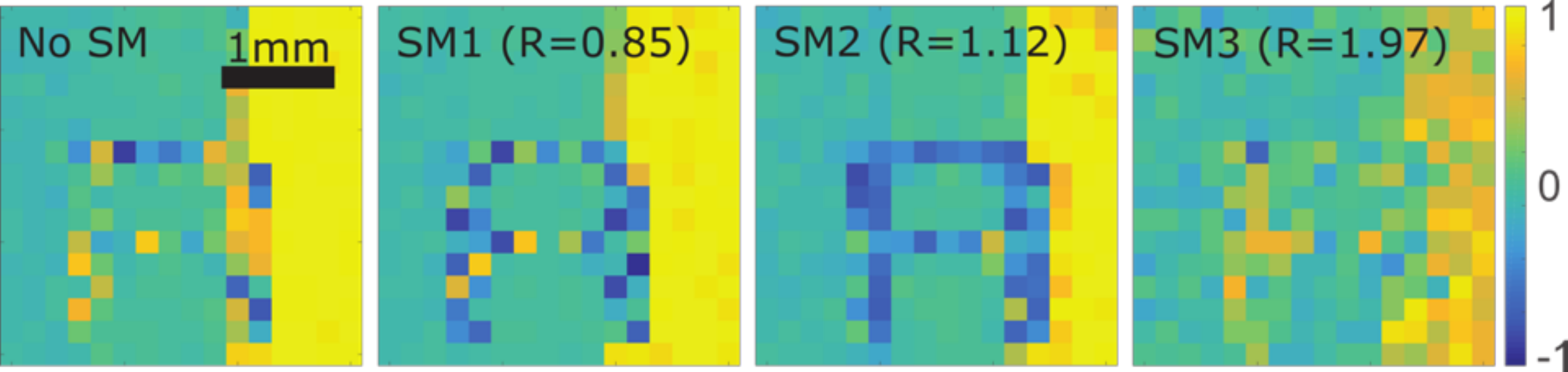}
	\caption{Normalised $M_{01}$ element of the Mueller matrix obtained for imaging through different scattering media.  }
	\label{fig:SNRimg}
\end{figure}

The difference between the matrix elements of the normalised Mueller matrices obtained with and without a scattering medium present was quantified by computing the root-mean-squared error (RMSE) for each normalised Mueller matrix element across all image pixels as shown in Figure \ref{fig:RMSE} for all three scattering media. Pixels related to the opaque letter R (found via thresholding the $M_{00}$ matrix element) consist primarily of noise that was further amplified upon normalisation, and were hence excluded when computing the RMSE. It can be seen that the average RMSE was $\approx 0.1$ for SM1 and SM2, but increased to $\approx 0.3$ for SM3. The increase in RMSE reflects the decrease in SNR at greater thicknesses resulting from more light being scattered out of the collection angle of the PSA, as well as the larger depolarisation. A significant decrease in signal level is evident from comparing the measured intensities across all three scattering media. For example, the total intensity transmitted through SM3 was 85$\%$ lower than that of SM1 for the first analysed and input polarisation state. Consequently, the reconstructed images were visibly noisier for thicker phantoms, as seen in Figure \ref{fig:SNRimg}.  Nevertheless, polarimetric information was still recoverable even for SM3. For instance, noting that the first row of the Mueller matrix of an ideal linear polariser with its transmission axis oriented in the $x$ direction is $ [1,1,0,0]$, whereas in comparison, for scotch tape and the glass substrate it is theoretically $[1,0,0,0]$, the right-hand region  of the object, corresponding to the linear polariser, can be clearly distinguished Figure \ref{fig:SNRimg}. The full Mueller matrix presented in the supplementary figures shows that all three materials in the test object can be well distinguished. 
Finally, although not reported here, imaging using the full spatial dependence of $\mathbf{M}_m^{\txtpow{SM}}$ gave comparable results~\cite{Seow2020}.

In summary, this work has demonstrated single pixel polarimetric imaging through scattering media for the first time. Using a proposed imaging model, it was shown that under coherent illumination, single pixel polarimetric imaging through scattering media was possible for pixel sizes larger than the spatial correlation length of the scattering medium for which contributions from different input pixels sum incoherently. This was further demonstrated in experiments in which the spatially resolved Mueller matrix of a test object hidden behind  scattering phantoms with thicknesses up to twice the TMFP was successfully reconstructed. As with most techniques, the imaging depth of single pixel polarimetric imaging is mainly limited by the decrease in SNR as the thickness of the scattering medium increases. Nevertheless, the utilisation of scattered has enabled imaging at greater depths than imaging with ballistic light alone. To achieve imaging resolutions closer to the correlation length of the scattering medium, smaller pixel sizes would be required. In this case, use of spatially incoherent illumination would be beneficial to ensure contributions from different input pixels sum incoherently. Possible future developments of this technique include an optimisation of the experimental setup to enhance the SNR and reduce the acquisition time. Finally, design of better reconstruction algorithms which are robust to noise would help improve imaging performance.
\noindent

\paragraph{Funding Information.} DSO National Laboratories (Singapore) and the Royal Society (UK)

\paragraph{Disclosures.} The authors declare no conflicts of interest.

\pagebreak
\appendix
%
\setcounter{figure}{0}
\renewcommand{\thefigure}{A\arabic{figure}}

\section{Supplementary Figures}
\begin{figure}[h!]
	\centering
	\includegraphics[width=0.98\linewidth]{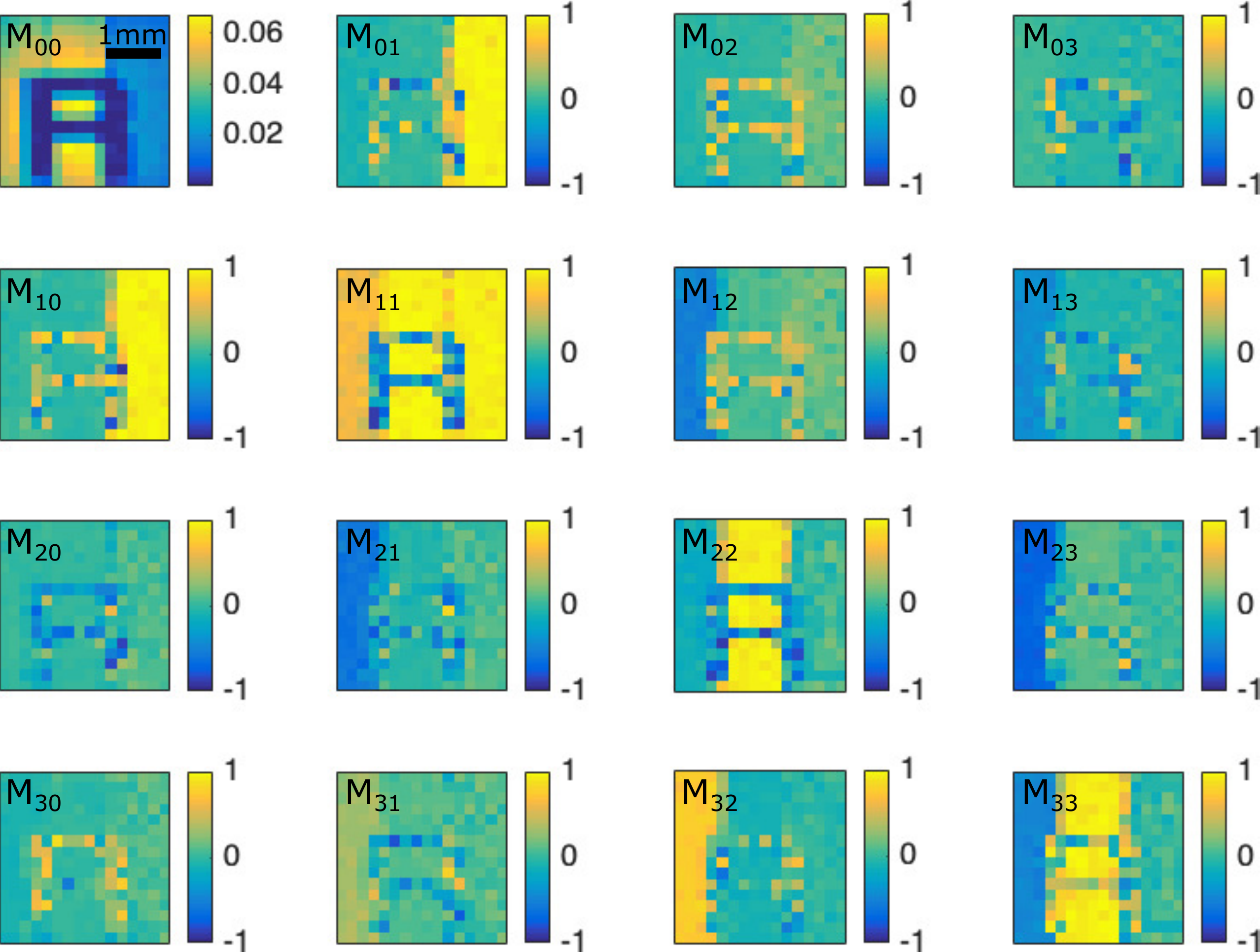}
	\caption{Spatially resolved Mueller matrix for the test object without SM1 present, with pixels in all matrix elements other than the $M_{00}$ element normalised to their respective $M_{00}$ values. }
\end{figure}

\begin{figure}[t!]
	\centering
	\includegraphics[width=0.98\linewidth]{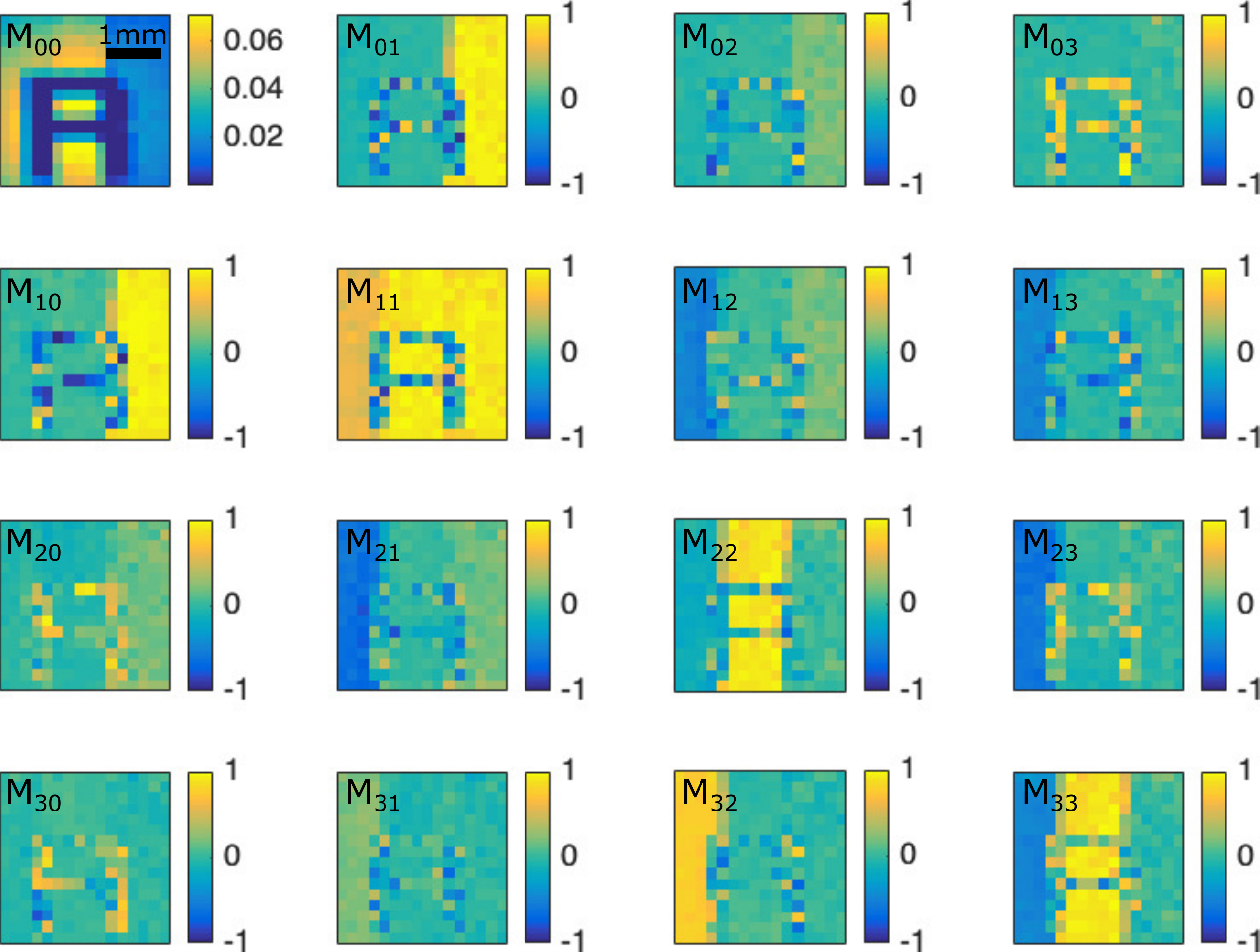}
	\caption{Spatially resolved Mueller matrix for the test object with SM1 present, with pixels in all matrix elements other than the $M_{00}$ element normalised to their respective $M_{00}$ values. }
\end{figure}
\noindent

\end{document}